\documentclass{article}

\usepackage[preprint]{neurips_2026}

\usepackage[utf8]{inputenc} 
\usepackage[T1]{fontenc}    
\usepackage{hyperref}       
\usepackage{url}            
\usepackage{booktabs}       
\usepackage{amsfonts}       
\usepackage{nicefrac}       
\usepackage{microtype}      
\usepackage{xcolor}         

\usepackage{multirow}       

\usepackage[pdftex]{graphicx}

\definecolor{winter}{rgb}{0.85,0.08,0.2}
\definecolor{summer}{rgb}{0.95,0.53,0.18}         
\definecolor{spring}{rgb}{0.02,0.93,0.68}
\definecolor{autumn}{rgb}{0.02,0.68,0.9}

\hypersetup{
    colorlinks=true,
    citecolor=autumn,
    linkcolor=magenta,
    filecolor=magenta,      
    urlcolor=cyan,
    }

\title{Multimodal Alignment and Preference Optimization for Zero-Shot Conditional RNA Generation}

\author{%
  Roman Klypa \\
  Univ. Grenoble Alpes, CNRS, Grenoble INP, LJK \\
  38000 Grenoble, France \\
  \texttt{Roman.Klypa@univ-grenoble-alpes.fr} \\
  \And
  Alberto Bietti \\
  Center for Computational Mathematics, Flatiron Institute \\
  162 5th Ave, New York, NY 10010, USA \\
  \texttt{Alberto.Bietti@gmail.com} \\
  \And
  Sergei Grudinin \\
  Univ. Grenoble Alpes, CNRS, Grenoble INP, LJK \\
  38000 Grenoble, France \\
  \texttt{Sergei.Grudinin@univ-grenoble-alpes.fr} \\
}

\begin{document}

\maketitle

\begin{abstract}
  The design of RNA molecules that interact with specific proteins is a critical challenge in experimental and computational biology. Despite recent progress in natural language modeling and deep learning-based protein design, there remains significant room to improve the frequency of successful interactions and the authenticity of generated sequences for functional applications. In this work, we frame conditional RNA sequence generation as a multi-stage alignment problem, introducing Moirain: a suite of models optimized via multimodal supervised fine-tuning (SFT) and Direct Preference Optimization (DPO). Our approach begins with large-scale pretraining on diverse RNA corpora to capture the fundamental grammars of sequence plausibility. To achieve target-specific generation, we employ a multimodal SFT architecture that conditions RNA synthesis on protein structural and sequential features. Finally, we leverage DPO to refine the model using synthetic interaction data: taking advantage of DPO’s unique ability to navigate non-aligned preference spaces, we improve functional fitness without collapsing the learned natural distribution. Extensive evaluation of the Moirain series (Moirain-Base, -Multi, and -DPO) demonstrates that our framework consistently produces novel, diverse, and biologically plausible RNA sequences with superior binding affinities compared to existing baselines. 
\end{abstract}

\section{Introduction}

Deep learning has transformed a broad spectrum of scientific and technical domains by enabling the modeling of complex, high-dimensional data. In structural biology, these methods have redefined the prediction and design of proteins, as evidenced by AlphaFold2 \cite{jumper_highly_2021}, Chroma \cite{ingraham_illuminating_2023}, and ESM3 \cite{hayes_simulating_2024}. In parallel with biological developments, Transformer-based \cite{vaswani_attention_2023} autoregressive \cite{bengio_neural_2000} language models have established a new state of the art for sequence generation. Modern Large Language Models (LLMs) now serve as powerful generative engines that excel across a broad range of specialized tasks, from document summarization~\cite{brown_language_2020} to complex reasoning~\cite{wei_chain--thought_2023}.

Despite broad progress across language modeling and structural biology, many areas involving biological sequences have yet to experience a comparable defining shift in performance and utility. In particular, the de novo generation of ribonucleic acid (RNA) sequences binding to specific protein targets has not yet reached the same level of maturity as that of protein engineering. This task is central to understanding the interactions between RNAs and proteins \cite{li_rna-protein_2024,fasogbon_recent_2025}, which govern essential biological processes such as gene regulation, splicing, and translation \cite{hentze_brave_2018}. A primary objective in this field is the design of aptamers: short, single-stranded RNA sequences capable of binding specific proteins with high affinity. Because these molecules can function as inhibitors, probes, or delivery agents, they offer versatile applications in therapeutics and diagnostics \cite{guo_engineering_2010, thavarajah_rna_2021}.

Several studies have explored RNA generation in this domain. More classical approaches exploited evolutionary signals and statistical models \cite{kim_computational_2007,kim_ragpools_2007,aita_biomolecular_2010,tseng_entropic_2011,zhang_single-step_2025}, molecular modeling \cite{torkamanian-afshar_silico_2021}, and Monte Carlo tree search \cite{lee_predicting_2021, wang_discrete_2022, shin_aptatrans_2023, obonyo_rna_2024}. Recent works used long short-term memory models \cite{im_generative_2019, park_discovering_2020}, conditional variation autoencoders  \cite{chen_generating_2022,iwano_generative_2022,andress_daptev_2023}, adversarial approach \cite{ozden_rnagen_2023}, diffusion processes \cite{wang_aptadiff_2024,zhang_rnagenesis_2024} and LLM fine-tuning \cite{zhao_generrna_2024}. Traditionally, RNA design has largely focused on specialized architectures requiring dedicated training for each individual protein. A more compelling challenge lies in the immediate synthesis of binders for novel targets. Recent research has increasingly adopted LLM-inspired solutions to address this zero-shot conditional task \cite{nori_rnaflow_2024,klypa_bang_2025,tabrizi_rnatranslator_2025,tabrizi_rnax_2025}. Notably, the most effective approaches remain at the sequence level, forgoing explicit structural modeling to avoid the inaccuracies inherent in current prediction tools \cite{rhiju_nucleic_2024}. Despite these developments, there remains clear space to improve the frequency of successful interactions and the authenticity of the generated sequences required for functional biological applications.

The primary objectives of this work are to design RNAs characterized by high binding affinity to the protein of interest and biological plausibility. We aim for the model to generalize across both the conditioning manifold and the output space, producing novel and diverse sequences that maintain high performance even for targets significantly different from those in the training set. To achieve this, we adopt a methodological framework consistent with the development of modern LLMs, comprising large-scale pretraining to capture general patterns \cite{radford_learning_2021}, followed by instruction tuning \cite{wei_finetuned_2022,chung_scaling_2022,raffel_exploring_2023}, also known as Supervised Fine-Tuning (SFT) \cite{ouyang_training_2022, bai_training_2022}, and reinforcement learning (RL) alignment to refine the focus of the model on specific objectives \cite{ziegler_fine-tuning_2020, ouyang_training_2022, rafailov_direct_2023}.

\begin{figure}
  \centering
  \includegraphics[width=\linewidth]{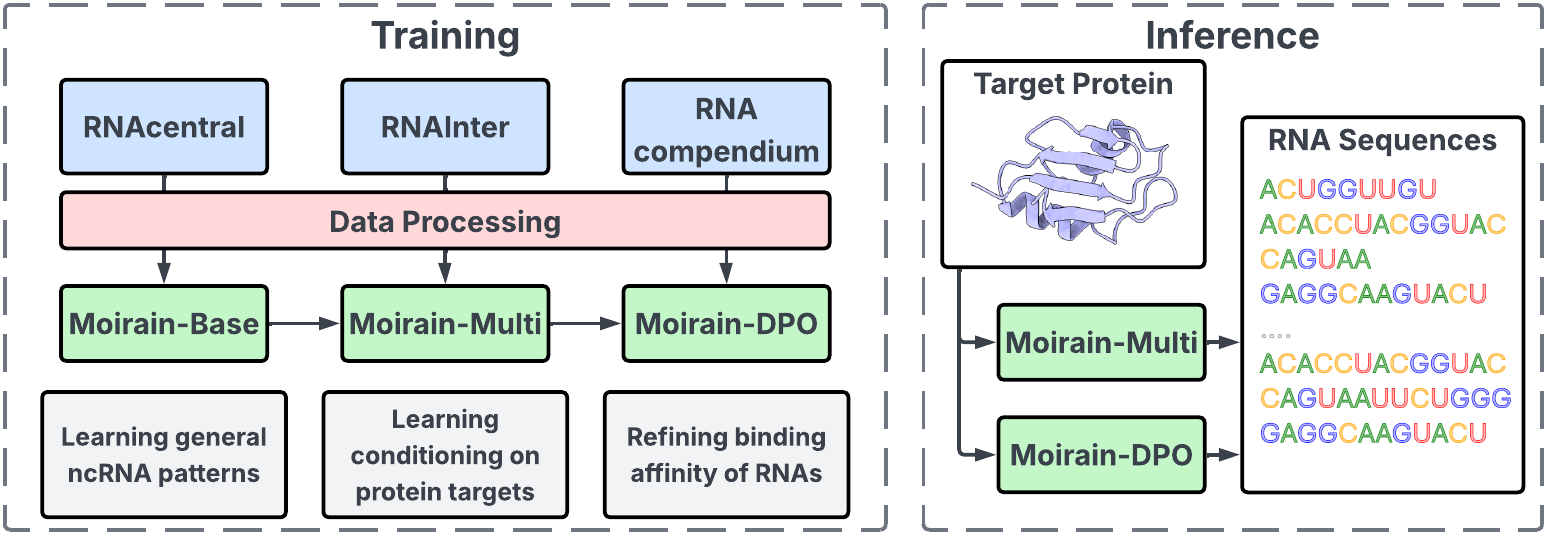}
  \caption{Overview of the Moirain framework. Schematic of the sequential training pipeline, comprising Moirain-Base, Moirain-Multi, and Moirain-DPO, alongside the inference workflow for zero-shot, protein-conditioned RNA generation.}
  \label{fig:pipeline}
\end{figure}

We begin with a pretraining stage on an extensive corpus of non-coding RNAs, establishing a broad biological context for the model. To enable protein-conditioned generation, we perform multimodal supervised fine-tuning on a dataset of targets and their cognate RNA partners, aligning the model’s output distribution with the subspace of interactive sequences. In a subsequent Direct Preference Optimization (DPO) \cite{rafailov_direct_2023} phase, we train our model on preference pairs of RNAs ranked by binding affinity, tuning it toward specific performance objectives rather than mere sequence likelihood. Ultimately, our method (illustrated in Fig. \ref{fig:pipeline}) is designed to achieve a high-performance, protein-conditioned design while preserving the essential biological authenticity acquired during the earlier training stages. Our main contributions can be summarized as follows:

\begin{enumerate}
    \item We frame target-conditioned RNA sequence design as a multimodal alignment problem, integrating LLM architectures with supervised fine-tuning and preference optimization.

    \item We develop Moirain-Base, a foundational RNA generative model, and extended it via a specialized multimodal SFT framework into Moirain-Multi, enabling zero-shot generation of novel biologically plausible sequences conditioned to bind to a specific protein target. We demonstrate that the choice of SFT loss function is a critical determinant of performance in the subsequent alignment stage.

    \item We curate a novel preference dataset and apply Direct Preference Optimization to Moirain-Multi, resulting in Moirain-DPO. By leveraging the DPO unalignment effect, we optimize for key binding motifs while filtering out synthetic artifacts, allowing us to achieve state-of-the-art performance in target interaction without compromising RNA sequence authenticity.
\end{enumerate}

\section{Theoretical Preliminaries and Backgrounds}

\subsection{Large Language Models}
Large Language Models are trained as next-token predictors over a discrete vocabulary $\mathcal{V}$. Given a sequence of tokens $x_{1:L} = (x_1,...,x_L)$, the model defines a conditional distribution $p_\theta( \cdot | x_{<l})$ over the next token $x_l$. Training consists of minimizing the Cross-Entropy (CE) between the predicted distribution and the data $\mathcal{D}$:
\begin{equation}
    \mathcal{L}_{\mathrm{CE}}(\theta) = - \mathbb{E}_{x \in \mathcal{D}} \left [ \log p_\theta(x_l | x_{<l}) \right ].
\end{equation}
By applying the chain rule, the joint probability of an entire sequence $x_{1:L}$ under the model parameters $\theta$ is decomposed into the product of the conditional distributions:
\begin{equation}
p_\theta(x_{1:L}) = \prod_{l=1}^{L} p_\theta(x_l | x_{<l}),
\end{equation}
thus making the training equivalent to maximizing the likelihood of the observed sequences.

\subsection{Supervised Fine-Tuning}

In Supervised Fine-Tuning, LLMs are adapted to specific tasks using sequences that combine a prompt and a response. The model conditions on the prompt as a fixed prefix but is optimized exclusively on the response. This ensures focus on generating the correct outputs for the task, while leveraging existing pretrained knowledge. This fine-tuning stage is often performed via the Low-Rank Adaptation (LoRA) \cite{hu_lora_2021} technique, which reduces the effective number of updated parameters, to improve efficiency and reduce overfitting as well as forgetting~\cite{biderman_lora_2024}.

The standard training objective for SFT is the Cross-Entropy loss. However, CE on relatively small datasets has been observed to reduce generative diversity \cite{omahony_attributing_2024,klypa_diversity_2026}, thus undermining downstream exploration and subsequent alignment. To address this issue, recent work has explored SFT variants employing modified loss functions to preserve output diversity, such as the Tempered Focal (TOFU) loss \cite{klypa_diversity_2026}, which targets both the forgetting of pretrained knowledge and the neglect of underrepresented samples in the fine-tuning dataset. To achieve this, TOFU reweights the Cross-Entropy gradients and applies temperature adjustment to the predicted distribution:

\begin{equation}
\mathcal{L}_{\mathrm{TOFU}}(\theta) = - \mathbb{E}_{x \in \mathcal{D}} \left [ \mathrm{sg} \left [ g(p_\theta,\gamma) \right ] \beta \log p^\beta_{\theta} \right ].
\end{equation}

In this formulation, focal term $g(p, \gamma) = (1-p)^\gamma - \gamma p (1-p)^{\gamma-1} \log p$ is detached from the gradient computation.

\subsection{Multimodality}

Multimodal large language models integrate visual and textual representations through varying architectural strategies. CLIP \cite{radford_learning_2021} established this field by using contrastive pretraining to create a shared embedding space. To incorporate frozen vision encoders, Flamingo \cite{alayrac_flamingo_2022} utilized cross-attention layers, whereas BLIP-2 \cite{li_blip-2_2023} introduced a lightweight query transformer to bridge the modality gap. Further simplifying this paradigm, LLaVA \cite{liu_visual_2023} projects visual features directly into the LLM input space, achieving alignment through instruction tuning on image–text pairs.

\subsection{Preference Optimization}

At its core, Preference Optimization is an extension of the reinforcement learning from human feedback (RLHF) framework \cite{ziegler_fine-tuning_2020}, which traditionally utilizes Proximal Policy Optimization (PPO) \cite{schulman_proximal_2017} to maximize a scalar reward signal under a Kullback–Leibler divergence constraint. While PPO serves as a general-purpose reinforcement learning algorithm capable of optimizing a policy against any arbitrary reward (human preferences or automated metrics), it requires maintaining multiple models and sampling from the policy during training. To alleviate these complexities, recent approaches derive closed-form expressions for the optimal policy, enabling direct optimization on preference data $\mathcal{D}$ containing samples $(x, y^+, y^-)$. Here, $y^+$ and $y^-$ represent the preferred and dispreferred completions for a given prompt x, respectively. Most of the Preference Optimization methods can be unified under a general objective:
\begin{equation}
\mathcal{L}_{\mathrm{PO}}(\theta) := \mathbb{E}_{(x, y^+, y^-) \sim \mathcal{D}} \left [ \ell_{x, y^+, y^-} \left ( \log p_\theta(y^+ \mid x) - \log p_\theta(y^- \mid x) \right ) \right ],
\end{equation}
where $\ell_{x, y^+, y^-} : \mathbb{R} \rightarrow \mathbb{R}^+$ is convex and differentiable.

The Direct Preference Optimization (DPO) objective \cite{rafailov_direct_2023} served as the seminal work in this area, establishing the paradigm of direct policy alignment by leveraging the analytical relationship between the reward and the policy. It employs a sigmoid function and a reference model $p_{\mathrm{ref}}$ to transform the preference task into a binary classification problem, resulting in the objective:
\begin{equation}
\mathcal{L}_{\mathrm{DPO}}(\theta) := \mathbb{E}_{(x, y^+, y^-) \sim \mathcal{D}} \left[- \log \sigma \left(\beta \left( \log \frac{p_\theta(y^+ \mid x)}{p_{\mathrm{ref}}(y^+ \mid x)} - \log \frac{p_\theta(y^- \mid x)}{p_{\mathrm{ref}}(y^- \mid x)} \right) \right) \right],
\end{equation}
where $\beta > 0$. Here, $p_{\mathrm{ref}}$ represents the model's state at the start of the DPO optimization, serving as a baseline to prevent $p_\theta$ from deviating too far. DPO has recently demonstrated success across diverse biological applications, ranging from protein design to the optimization of genomic sequences \cite{nguyen_sequence_2024,heinzinger_teaching_2025,listgarten_how_2026}.

\section{Proposed Method}
\subsection{Base Model Pretraining}

The primary goal of our pretraining is to capture general RNA patterns. This foundation ensures the authenticity of generated sequences, which is the primary determinant of whether they remain chemically viable and functional within a complex cellular environment. However, one must distinguish between two evolutionary ``languages'': coding RNAs, constrained by the genetic code for protein synthesis, and non-coding RNAs (ncRNAs), which prioritize direct cellular function over information transfer. Since our objective is to design binding partners, we focus exclusively on ncRNAs, as they constitute the vast majority of functional RNA–protein interactions. Additionally, not being translated, these molecules have a higher likelihood of remaining intact and available to reach their intended targets.

Guided by these considerations, we utilized RNAcentral \cite{thernacentralconsortium_rnacentral_2019}, the largest available database of non-coding RNAs, for the pretraining stage. To ensure data quality and reduce redundancy, we deduplicated the raw sequences, resulting in a final training set of 16.6 million unique RNAs. Naturally, RNA is composed of a four-letter alphabet. To enable a larger context window, we tokenized the sequences with Byte Pair Encoding (BPE) \cite{sennrich_neural_2016}. This approach effectively compresses the data by reducing RNA length. We fix the vocabulary size to 256 to limit the long low-frequency tail and ensure sufficient training signal per embedding. The resulting train corpus size is 3.2 billion tokens.

Our base model, Moirain-Base, comprises 302 million parameters and follows the LLaMA \cite{touvron_llama_2023} architecture family, which we optimized for stable training. We trained it for one epoch \cite{muennighoff_scaling_2023}; additional data and training details are provided in the Appendix~\ref{app:pretrain}.

\subsection{Multimodal Supervised Fine-Tuning}

To adapt Moirain-Base for conditional RNA generation, we perform multimodal supervised fine-tuning using paired interaction data from RNAinter \cite{kang_rnainter_2022}, comprising 203,811 pairs after processing. This approach is analogous to instruction tuning: the protein serves as a functional prompt that dictates the generated RNA response. However, because proteins represent a different semantic modality from the RNA sequences used during pretraining, our SFT process is specifically designed to bridge these distinct biological spaces, enabling the model to condition its generation on cross-modal structural and sequential information. The protein structure represents the three-dimensional spatial arrangement of a molecule, commonly referred to as its fold, that directly dictates its biological function, including its capacity for specific interactions. Consequently, its inclusion as input features serves the purpose of grounding RNA generation in the physical reality of the target, while bypassing a massive computational burden (as evidenced by the complexity of AlphaFold, for example) of learning theoretically possible sequence-to-fold mapping from scratch.

Due to the aforementioned reasons, we found projection-based conditioning to be unsuitable in our case. Instead, we adopted a cross-attention framework inspired by Flamingo and BLIP-2 to integrate protein information. To ensure parameter efficiency, we deviated from the standard Flamingo architecture by omitting the training of new feed-forward layers. This decision was based on the observation that the SFT dataset does not introduce novel sequential patterns beyond those already captured during RNA pretraining. Consequently, we applied Low-Rank Adaptation to the pretrained weights, as it provides a natural mechanism to constrain the generative space and condition the model on external features without disrupting the underlying RNA representations \cite{biderman_lora_2024}. 

\begin{figure}[t]
  \centering
  \includegraphics[width=\linewidth]{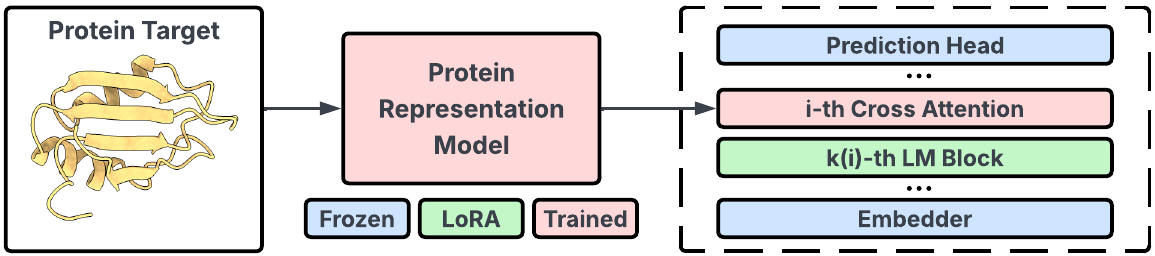}
  \caption{Schematic of the Moirain-Multi Cross-Attention Architecture training pipeline. The framework integrates protein features via $M$ cross-attention blocks ($M<N$) uniformly distributed across the $N$-layer stack according to Equation~\ref{eq:crosspos}. Parameter efficiency is maintained by utilizing LoRA of foundational RNA weights instead of adding fully trainable feed-forward layers.}
  \label{fig:cross}
\end{figure}

To maintain consistency with our previous architectural optimizations, we reduced the number of cross-attention blocks to $M<N$, where $N$ denotes the total count of original blocks. These $M$ modules are distributed uniformly throughout the stack, with the first and last ones positioned immediately following the 1st and N-th original layers, respectively. Under these constraints, the specific index $k(i)$ for each cross-attention block is defined by:
\begin{equation}
\label{eq:crosspos}
    k(i) = \textup{round} \left ( 1 + \frac{(i-1)(N-1)}{M-1} \right ), \quad i = 1, \dots, M.
\end{equation}
Such modules distribution ensures a consistent injection of conditional features across all levels of the model’s hierarchy, preventing signal degradation. The resulting architecture is schematically illustrated in Figure \ref{fig:cross}.

Having established the mechanism for integrating the protein modality, the next step is to define its source and encoding. We represent proteins by their sequences, structures from the AlphaFold Protein Structure Database (AFDB) \cite{fleming_alphafold_2025}, and per-residue pLDDT scores. To encode this information, we adopted the protein module from RNA-BAnG \cite{klypa_bang_2025-1}, which integrates transformer layers with geometric Invariant Point Attention (IPA) \cite{jumper_highly_2021}.

Integrating the multimodal architectural design described above into Moirain-Base results in the Moirain-Multi model, containing 331 million parameters, with 29 million being trainable and the remainder frozen. We fine-tuned it using two separate loss functions: standard Cross-Entropy as a baseline for sequence modeling and TOFU to enhance output diversity and mitigate overconfidence. Under both objectives, we trained the model for three epochs \cite{ouyang_training_2022,zhou_lima_2023}; additional data and training details are provided in Appendix~\ref{app:sft}.

\subsection{Preference Optimization}

In many cases, RNA binding affinity to the target is determined by only a fraction of the its total sequence, often confined to relatively small motifs \cite{ray_compendium_2013}. The intuitive logic is that, as multi-functional entities, RNA molecules must reserve nucleotide budget for tasks beyond protein interaction. Consequently, training without additional conditioning signals while aiming for high-performance generation is equivalent to solving a needle-in-a-haystack problem. Unfortunately, deep neural networks in general \cite{geirhos_shortcut_2020} and autoregressive modeling in particular \cite{dziri_faith_2023} have been observed to struggle in this regime as models tend to prioritize global patterns over short localized subsequences. This specific limitation has been previously noted in the context of RNA-protein interactions \cite{klypa_bang_2025-1}.

We address this challenge by employing a phase typically aimed at improving LLM performance: alignment via Preference Optimization. Unlike standard training, PO algorithms require a dataset structure where each prompt is associated with multiple distinct candidate responses to be evaluated or ranked. For our task, this format is provided by the RNA Compendium \cite{ray_compendium_2013}, which contains approximately 200,000 RNA sequences experimentally scored against roughly 240 protein targets. A critical feature of this dataset is that the sequences are synthetic and largely randomized. Consequently, using them as positive examples for supervised fine-tuning would be counterproductive, as we wish the model to maintain the natural RNA distribution learned during its previous training phases. Similarly, training a reward model on such data is risky, as it could bias the model to favor synthetic artifacts over the established biological priors.

Given the nature of the data and the task requirements, Preference Optimization via a pairwise ranking loss, such as Direct Preference Optimization, emerges as the currently most suitable strategy. The DPO objective is specifically designed to shift the probability distribution by increasing the likelihood gap between preferred and dispreferred examples rather than simply maximizing the probability of the former. Crucially, the probability of a preferred response has been observed to actually decrease during optimization \cite{pal_smaug_2024}. Although potentially prone to unintentional unalignment \cite{razin_unintentional_2025}, this effect serves our goal of bypassing synthetic noise. It allows the model to prioritize functional binding motifs without drifting from the original biological plausibility.

To implement this approach, we have constructed a preference dataset of 1000 pairs per 213 proteins and trained Moirain-Multi on it with LoRA (7.8 million active parameters) for 2 epochs \cite{ouyang_training_2022,zhou_lima_2023}, resulting in Moirain-DPO. Additional data and training details are provided in  Appendix~\ref{app:dpo}.

\begin{figure}
  \centering
  \includegraphics[width=\linewidth]{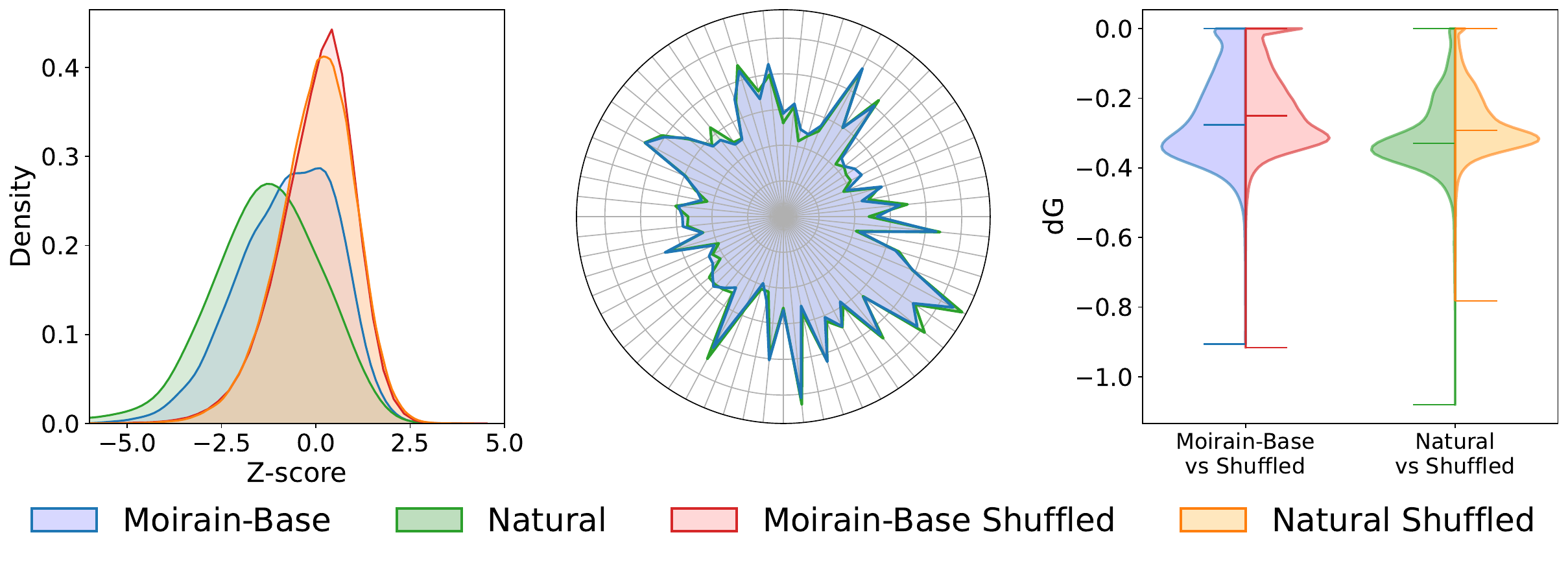}
  \caption{The comparison between generated (Moirain-Base) and natural RNA sequences. \textbf{(Left)} Z-score distributions for generated and natural sequences alongside their shuffled counterparts. \textbf{(Middle)} Radar plot of 3-mer frequency distributions (Fidelity $=0.022$), where each angular axis represents a distinct 3-mer. \textbf{(Right)} Distribution of dG shown via split-violin plots. For both the generated (left) and natural (right) groups, the left half of the violin represents the original sequences and the right half represents the shuffled controls. Horizontal markers indicate the median, minimum, and maximum values.}
  \label{fig:plaus_full}
\end{figure}

\section{Evaluation Setup}

As a baseline measure of biological plausibility, we compare the local patterns of generated sequences with those of natural ncRNAs. For this purpose, we employ the Total Variation distance between their respective 3-mer distributions, a metric we designate as {\em Fidelity}. Another, more convoluted and computationally demanding evaluation is based on the observation that non-coding RNAs typically exhibit a significantly lower Minimum Free Energy (MFE) than random sequences of the same dinucleotide composition \cite{freyhult_comparison_2005}. To quantify this phenomenon, we employ two distinct metrics from \cite{freyhult_comparison_2005}: the length-normalized MFE (dG), and the Z-score, which measures the deviation of a sequence's MFE from the mean of its shuffled counterparts. By comparing the distributions of these quantities between the generated and natural sequence sets, we can evaluate the model's ability to capture native-like foldability.

A generative model must demonstrate the ability to produce novel sequences beyond its training data while avoiding mode collapse. To ensure persistence of those qualities, we measure them per target. \textit{Novelty} is defined as the proportion of generated sequences with no detectable hits against a reference RNA set of choice, serving as a primary indicator for memorization detection. To isolate the impact of different training stages, we calculate this metric independently against each corresponding dataset used. \textit{Diversity} is measured as the proportion of cluster representatives at fixed similarity threshold, quantifying the breadth of the generated sequence space. The details of tools and parameters used can be found in Appendix~\ref{app:metrics}.

The utility of our model depends on its ability to generate sequences with high binding affinity. To evaluate it, we adopt the RNA-BAnG benchmark \cite{klypa_bang_2025-1}, comprising 71 proteins from the RNACompendium \cite{ray_compendium_2013}, each associated with a target-specific DeepCLIP model \cite{gronning_deepclip_2020}. Widely used for the task \cite{im_generative_2019,zhao_generrna_2024,klypa_bang_2025-1,tabrizi_rnatranslator_2025}, DeepCLIP is an interpretable lightweight CNN-based architecture for scanning probabilistic motifs, excelling when interactions are driven by sequential rather than structural RNA elements. To align with molecular design objectives, where the yield of high-performing candidates takes precedence over population averages, we introduce two threshold-based metrics: Hits$_x$ and Cov$_x$. Hits$_x$ denotes the fraction of generated sequences exceeding a binding score of $x$. Analogous to the pass@k metric in LLM evaluation, it quantifies the model's efficiency in producing high-affinity binders. Complementing this, Cov$_x$ measures the breadth of success across the target space, defined as the number of proteins for which at least $x$ percent of generated sequences surpass a binding probability of 0.7 \cite{klypa_bang_2025-1}.

\begin{table}
  \caption{Binding affinity and authenticity benchmarking results for tested models. Performance is quantified by: (i) Hits$_x$ (0–1), representing target-specific binding success; (ii) Cov$_x$ (0–1), denoting the coverage of the target manifold; and (iii) Fidelity (0–1), measuring the proximity to natural interacting RNA.}
  \label{tab:main}
  \centering
  \sc
  \small
  \setlength{\tabcolsep}{5pt}
  \begin{tabular}{llcccccc}
    \toprule
    Method & Loss & Hits$_{0.5}$$\uparrow$ & Hits$_{0.7}$$\uparrow$ & Hits$_{0.9}$$\uparrow$ & Cov$_{0.25}$$\uparrow$ & Cov$_{0.5}$$\uparrow$ & Fidelity $\downarrow$ \\
    \midrule
    RNATranslator & & 0.45 & 0.36 & 0.25 & 0.56 & 0.31 & 0.18 \\
    RNA-BAnG & & 0.57 & 0.51 & 0.42 & 0.77 & 0.48 & 0.23 \\
    \midrule
    \multirow{2}{*}{Moirain-Multi} & CE & 0.40 & 0.32 & 0.22 & 0.69 & 0.13 & \textbf{0.05} \\
     & TOFU & 0.41 & 0.33 & 0.22 & 0.72 & 0.14 & \textbf{0.05} \\
    \multirow{2}{*}{Moirain-DPO} & CE & 0.58 & 0.51 & 0.39 & 0.77 & 0.58 & 0.12 \\
     & TOFU & \textbf{0.68} & \textbf{0.63} & \textbf{0.54} & \textbf{0.82} & \textbf{0.59} & 0.17 \\
    \bottomrule
  \end{tabular}
\end{table}

\section{Results}

Our primary objective for Moirain-Base is to determine whether it successfully replicates the structural and sequence characteristics of natural non-coding RNAs. As illustrated in Figure~\ref{fig:plaus_full}, the model achieves high Fidelity to natural 3-mer profiles. While the generated sequences appear to be less structured on average than natural ones (higher dG values), they nonetheless maintain a clear foldability signal, as evidenced by their Z-score distributions. Conversely, dG yields fewer insights, as the distributions for natural and generated sequences closely overlap with each other and with their respective shuffled baselines. This observation aligns with \cite{freyhult_comparison_2005}, which notes that dG varies significantly by ncRNA class and often lacks discriminative power when evaluating the heterogeneous mixtures present in our combined datasets.

For the task of conditional generation, we evaluate our models against existing open-weight baselines. To maintain a fair and focused comparison, we limit our scope to models capable of zero-shot generation \cite{nori_rnaflow_2024,klypa_bang_2025,tabrizi_rnatranslator_2025}, as many other existing methods are not directly applicable without extensive target-specific training. While RNAFlow is a notable existing model, prior benchmarks in the RNA-BAnG study demonstrated its limited efficacy in this specific task, therefore, we omit it from our evaluation to prioritize more competitive, high-performing methods. The results of the comparison are summarized in Table~\ref{tab:main}. The underwhelming performance of Moirain-Multi reinforces the hypothesis that a basic autoregressive approach is insufficient for this task. In contrast, Moirain-DPO achieves the best binding affinity across our tests. Crucially, incorporating TOFU loss during the SFT stage leads to substantial quality improvements. This suggests that a "relaxed" distribution is significantly easier to optimize, as it maintains the necessary flexibility for the model to adapt during preference tuning. While Fidelity decreases following the DPO stage, it remains superior to that of the alternative methods. 

A more focused analysis reveals that when benchmarking is narrowed to protein targets with low sequence similarity to both Moirain and RNA-BAnG training sets, Moirain-DPO exhibits a more pronounced performance degradation than its counterpart (detailed numerical results are provided in Appendix~\ref{app:results}). This suggests that while our approach excels in optimized interactions, generalizing to entirely unseen protein manifolds remains a standing challenge. Furthermore, it should be noted that the proteins in our benchmark are represented within the RNATranslator training data, precluding a direct assessment of that model's generalizability in this specific context.

All Moirain variants maintain a high degree of generational breadth. As anticipated, the TOFU loss demonstrates superior expressiveness compared to standard Cross-Entropy following the SFT stage. Notably, the peak performance of Moirain-DPO (when initialized via TOFU Moirain-Multi) coincides with its highest diversity and novelty scores, effectively ruling out the possibility of mode collapse. While outputs from Moirain-Base exhibit limited novelty, aligning with previous observations for RNA language models \cite{zhao_generrna_2024}, this metric improves significantly after fine-tuning. Both Moirain-Multi and Moirain-DPO produce highly original sequences. Finally, the absence of matches within synthetic databases validates our hypothesis regarding the suitability of DPO for this task.

\begin{table}
  \caption{Novelty and Diversity benchmarking across the Moirain suite. Metrics are reported on a scale of 0–1. For Moirain-Base, values represent global averages over the entire generated corpus; for Moirain-Multi and Moirain-DPO, values are calculated per-protein. Error bars denote the standard deviation. Novelty is assessed relative to the specific training RNA data of each respective stage.}
  \label{tab:novelty}
  \centering
  \sc
  \small
  \setlength{\tabcolsep}{4pt}
  \begin{tabular}{llcccc}
    \toprule
    Method & Loss & Novelty Base $\uparrow$ & Novelty Multi $\uparrow$ & Novelty DPO $\uparrow$ & Diversity $\uparrow$ \\
    \midrule
    Moirain-Base & & 0.42 & - & - & 0.73 \\
    \multirow{2}{*}{Moirain-Multi} & CE & $0.74_{\pm0.06}$ & $0.73_{\pm0.06}$ & - & $0.68_{\pm0.16}$ \\
     & TOFU & $0.87_{\pm0.03}$ & $0.86_{\pm0.03}$ & - & $0.93_{\pm0.03}$ \\
    \multirow{2}{*}{Moirain-DPO} & CE & $0.85_{\pm0.06}$ & $0.83_{\pm0.07}$ & \textbf{1.0} & $0.91_{\pm0.03}$ \\
     & TOFU & $\textbf{0.98}_{\pm0.01}$ & $\textbf{0.98}_{\pm0.01}$ & \textbf{1.0} & $\textbf{0.98}_{\pm0.02}$ \\
    \bottomrule
  \end{tabular}
\end{table}

\section{Conclusion}

In this work, we addressed the problem of designing RNA molecules conditioned on specific protein targets, prioritizing both binding affinity and biological plausibility. We framed this challenge as a multimodal alignment problem, integrating Large Language Model architectures with supervised fine-tuning and preference optimization. We first developed Moirain-Base, a foundational generative model, and extended it through a specialized multimodal SFT framework into Moirain-Multi, enabling the zero-shot synthesis of novel, authentic sequences tailored to protein targets. We hypothesized that a purely autoregressive approach would struggle with the needle-in-a-haystack nature of binding sites. Our results validate this assumption, demonstrating that an alignment stage is essential for reliably navigating the space of high-affinity binders.

We curated a novel preference dataset and applied Direct Preference Optimization to Moirain-Multi, resulting in Moirain-DPO. By leveraging the DPO unalignment effect, we successfully optimized for key binding motifs while filtering out artifacts inherent in synthetic training data. This approach allowed us to achieve state-of-the-art target interaction without compromising sequence plausibility. The absence of overlap between our generated sequences and the synthetic DPO training set, combined with authenticity scores that surpass baseline methods, confirms the suitability of the DPO framework for navigating the complex trade-offs of conditional RNA design. Crucially,  we demonstrate that the integration of the TOFU objective during the SFT stage is a primary driver of our model's performance. The resulting improvements in binding affinity, diversity, and novelty underscore the critical role of loss function selection in the alignment pipeline.

While Moirain-DPO sets a new benchmark, the primary challenge remains achieving robust generalization across novel protein targets. We also recognize that further refining biological plausibility offers a promising optimization avenue. Ultimately, as a natural extension of our in silico evaluation, wet-lab validation is the necessary step to confirm the therapeutic potential of the generated sequences.

In summary, our work demonstrates that the integration of previously untapped data sources with a dedicated multimodal architecture provides a powerful framework for navigating complex biological constraints. Although challenges remain, our approach to aligning optimization objectives and refinement stages offers a promising direction for RNA design, achieving superior performance. We hope the methodologies and insights presented here inspire further research into the grand challenges of therapeutic design.

\paragraph{Software and Data.}
The code and the models, along with the models weights, will be available upon publication.

\section*{Impact Statement}

The main purpose of this work is to advance the field of generative models. However, the applications of this method may have social and industrial benefits. Potential applications include in-silico SELEX approaches, RNA vaccine design, the development of novel drugs, and some other therapeutic tasks.

\begin{ack}
This work was performed using HPC resources from GENCI–IDRIS (Grant 2025-AD011015647R1). This work has benefited from state aid managed by the National Research Agency under the France 2030 program (Grant ANR-23-IACL-0006).  
\end{ack}

\bibliographystyle{unsrt}
\bibliography{main}

\newpage
\section*{Appendix}
\appendix

\counterwithin{figure}{section}
\counterwithin{table}{section}
\counterwithin{equation}{section}

\renewcommand{\thefigure}{\Alph{section}.\arabic{figure}}
\renewcommand{\thetable}{\Alph{section}.\arabic{table}}
\renewcommand{\theequation}{\Alph{section}.\arabic{equation}}

\section{Technical appendices and supplementary material}

\subsection{Pretraining Details}
\label{app:pretrain}

\paragraph{Data Curation \& Preprocessing} The pre-training corpus was sourced from RNAcentral. To ensure sequence integrity and computational efficiency, we excluded any samples containing non-standard nucleotides or those with lengths exceeding 20,000 or falling below 16 nucleotides. The remaining dataset was clustered using MMseqs2 (linclust) \cite{steinegger_clustering_2018} with a minimum sequence identity of 0.8 and a coverage threshold of 0.8. Cluster representatives were designated for the training set, with 10,000 randomly selected reserved as a held-out validation set.

\begin{figure}[h]
  \centering
  \includegraphics[width=\linewidth]{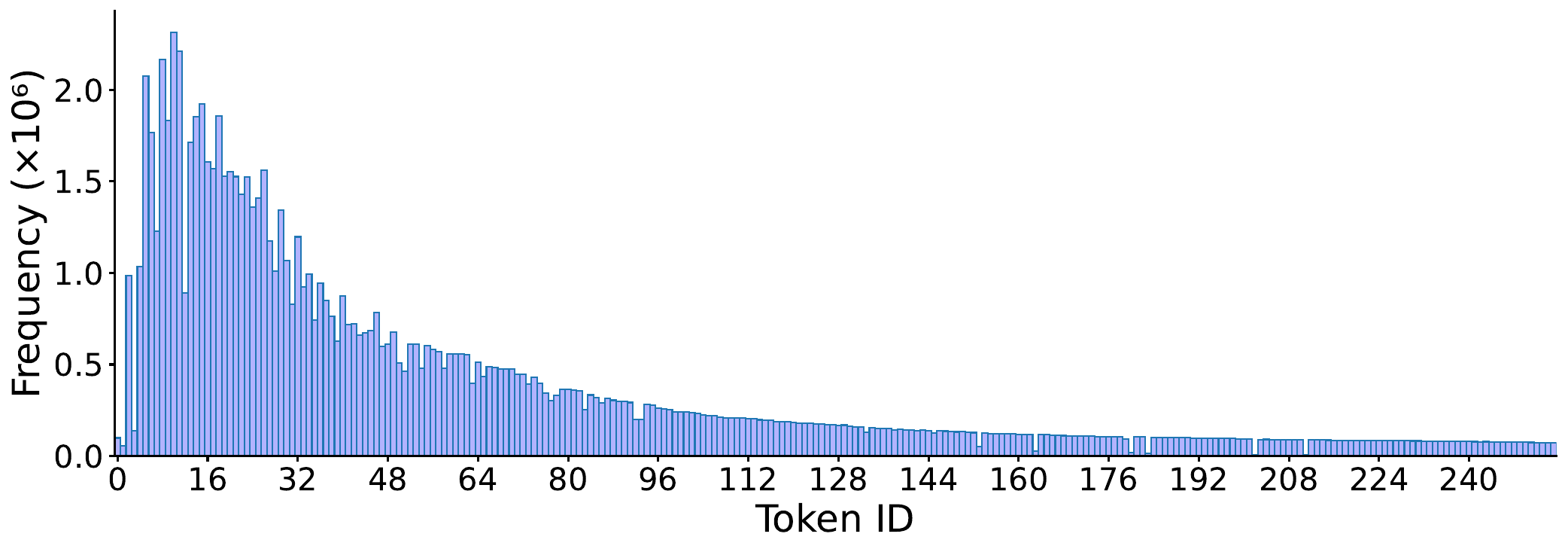}
  \caption{The plot illustrates the frequency of tokens within the data subset used for tokenizer training. Tokens are ordered along the x-axis by their respective token IDs, which corresponds to their BPE merge order.}
  \label{fig:vocab_freqs}
\end{figure}

\paragraph{Tokenization} We utilized a Byte Pair Encoding tokenizer. The vocabulary was learned from a subset of 500,000 sequences randomly sampled from the entire processed dataset. The frequencies of the resulting tokens on the same subset are depicted in Figure~\ref{fig:vocab_freqs}.

\paragraph{Architecture} The model consists of $N=24$ transformer blocks. Each block comprises self-attention layers and feed-forward networks utilizing GeLU activations \cite{hendrycks_gaussian_2016}. To ensure stable training at scale, we employed RMSNorm \cite{zhang_root_2019} for pre-attention normalization \cite{dehghani_scaling_2023}. Relative positional information was incorporated using Rotary Positional Encodings (RoPE) \cite{su_roformer_2023}, facilitating the handling of variable sequence lengths. Moirain-Base features a latent dimension of 1024, utilizing 16 attention heads with a hidden dimension of 64 each, and a feed-forward expansion factor (transition factor) of 4.

\paragraph{Training Configuration} Moirain-Base was trained on eight NVIDIA A100 GPUs with a global batch size of 64. We utilized a maximum context length of 2,048 tokens; for sequences exceeding this limit, random crops were sampled.

\paragraph{Optimization} Optimization was performed using the AMSGrad \cite{reddi_convergence_2019} variant of the Adam \cite{kingma_adam_2017} optimizer with default hyperparameters $(\beta_1 = 0.9, \beta_2 = 0.999)$. The learning rate was set to $5 \times 10^{-5}$, governed by a schedule consisting of a linear warmup over the first 10,000 steps, followed by a sinusoidal decay for the remainder of the training duration.

\subsection{Multi-Modal SFT Details}
\label{app:sft}

\paragraph{Data Collection \& Mapping} Interaction data was sourced from RNAInter, which provides protein-RNA pairs via database identifiers. Protein IDs were mapped to their corresponding UniProt \cite{the_uniprot_consortium_uniprot_2025} sequences, ensuring, where possible, that the UniProt gene nomenclature aligned with RNAInter records. For RNA, we restricted our selection to samples originating from NCBI \cite{brown_gene_2015} or miRBase \cite{kozomara_mirbase_2019} to ensure high-quality transcript mapping. These were mapped to RefSeq \cite{oleary_reference_2016}, filtered for non-coding RNA transcripts, and RNAcentral respectively. Any identifiers mapping to more than four distinct sequences were excluded to maintain data integrity. Successfully mapped proteins were grouped into FoldSeek \cite{van_kempen_fast_2024} clusters, while RNA sequences were clustered using the same parameters as the pre-training data (MMseqs2, 0.8 identity, 0.8 coverage).

\paragraph{Data Curation \& Preprocessing} To prevent the model from collapsing into a "one-size-fits-all" generation strategy, we implemented a filtering pipeline to address highly promiscuous interactors. We observed that less than 1\% of RNAs accounted for approximately 80\% of protein cluster interactions. To mitigate this:
\begin{enumerate}
    \item We removed RNA clusters that individually interacted with more than 40\% of all protein clusters.
    \item For each unique protein, we retained interactions only with RNA clusters that cover fewer than 64 protein clusters.
    \item In cases where a protein had no such partners, we preserved only two of its paired RNA clusters, with the least number of interactions with distinct protein clusters .
\end{enumerate}
To optimize computational throughput, we excluded any protein sequences longer than 512 amino acids and RNA sequences exceeding 512 tokens. The RNA-BAnG benchmark includes a "zero-similarity" subset, which contains proteins that share no detectable sequence similarity (via BLASTp \cite{altschul_basic_1990} with default parameters) with its own training data. It serves as a rigorous test for generalization across unseen target space. To maintain this independence in our own training, we withheld all protein clusters containing sequences with 50\% or greater similarity to this subset, also using BLASTp with default parameters. The final processed training set comprised 203,811 interaction samples, spanning 10,648 protein clusters and 12,287 RNA clusters.

\paragraph{Architecture} For the protein representation block, we utilize its default configuration from RNA-BAnG. The cross-attention architecture comprises $M=4$ blocks with a hidden dimension of 64 and 16 heads. Each block applies pre-normalization to both keys and queries. Positional encodings are omitted, as the cross-attention operates across different modalities. The pLDDT scores representing protein structure confidence are encoded by discretizing the values into 32 equal bins.

\paragraph{Training Configuration} TOFU loss parameters were set to recommended $\beta=0.8,\gamma=0.3$. Moirain-Multi was trained on four NVIDIA A100 GPUs with a global batch size of 32. During training, we sampled pairs sharing the same protein and RNA clusters at a rate of two pairs per epoch. We applied LoRA $(r=32,\alpha=32)$ to the keys, queries, projection, and feed-forward layers. 

\paragraph{Optimization} While the general optimizer architecture remained consistent with the pre-training phase, the learning rate was adjusted to $10^{-4}$ and the linear warmup period was shortened to 1,000 steps.

\subsection{Preference Optimization Details}
\label{app:dpo}

\paragraph{Data Curation} The preference dataset was constructed using experimental binding scores from the RNA Compendium. To create high-contrast preference pairs, we selected the 1,000 sequences with the highest binding scores and paired them with the 1,000 sequences possessing the lowest scores. To maintain the integrity of our generalization benchmarks, we excluded any proteins belonging to MMseqs2 clusters that shared 40\% or greater sequence similarity with the "zero-similarity" test subset. Furthermore, we restricted the dataset to proteins with lengths below 512 amino acids. The resulting training set comprised 213,000 preference samples.

\paragraph{Training \& Optimization} The DPO objective was optimized with a default $\beta=0.1$ coefficient. We applied LoRA $(r=16,\alpha=16)$ to the keys, queries, projection, and feed-forward layers. Moirain-DPO was trained on four NVIDIA A100 GPUs with a global batch size of 16. We utilized the same optimizer configuration as the pre-training phase, with the exception of the linear warmup, which was adjusted to 5,000 step.

\section{Inference Details}

During the generation of RNA sequences, we employed a temperature of $T=1$ across all models. For Moirain-Base, we utilized full random sampling to explore the learned sequence space, performing unconstrained generation of 10,000 sequences with a maximum length of 512 tokens. For our tuned variants, Moirain-Multi and Moirain-DPO, we implemented Nucleus Sampling (Top-p) \cite{holtzman_curious_2020} with a threshold of $p=0.9$ to balance diversity and coherence. In conditioned generation tasks, we generated 1000 sequences per target protein, with the decoding process constrained to a maximum of 50 tokens.

\section{Metrics}
\label{app:metrics}

\paragraph{dG \& Z-score} We used uShuffle \cite{jiang_ushuffle_2008} to create 100 distorted versions of each sequence via dinucleotide-preserving permutations. These were used to compare the dG of the original sequences against a shuffled background and to calculate MFE Z-scores. The MFE was computed using RNAfold \cite{lorenz_viennarna_2011}, where we included only sequences shorter than 256 nucleotides to maintain adequate computational time. The resulting analysis was conducted on 6,869 generated sequences and 3,272 natural sequences.

\paragraph{Fidelity} To evaluate Fidelity, we compared the 3-mer distributions of generated sequences against specific reference sets. For Moirain-Base, the comparison was performed against a held-out set of natural ncRNAs. For the remaining models, comparisons were made against the interacting RNA sequences from the multi-modal SFT training set. We restricted the analysis to sequences shorter than 512 nucleotides for the unconditional task and shorter than 64 nucleotides for conditional tasks to ensure that potentially cropped sequences were excluded from the evaluation. The resulting comparison sets comprise 8,956 samples for Moirain-Base, 6,756 natural ncRNAs, and 2,726 interacting RNAs, with approximately 60,000 to 70,000 samples for each conditioned model variant.

\paragraph{Novelty \& Diversity} For the calculation of Novelty, we employed MMseqs2 search \cite{steinegger_mmseqs2_2017} with the search type parameter set to 3. Searches were performed individually against the respective training sets for Moirain-Base and Moirain-Multi, and against the preferred examples within the Moirain-DPO training set. To determine Diversity clusters, we utilized MMseqs2 (linclust) with a 0.8 similarity threshold and a 0.8 coverage.

\paragraph{Binding Affinity} To compute binding affinity scores, we cropped all sequences to 50 nucleotides and excluded those with a length of less than 6 nucleotides. These constraints were applied to adhere to DeepCLIP size restrictions and because very short sequences lack biological relevance in this context.

\section{Additional Results}
\label{app:results}

\begin{figure}
  \centering
  \includegraphics[width=\linewidth]{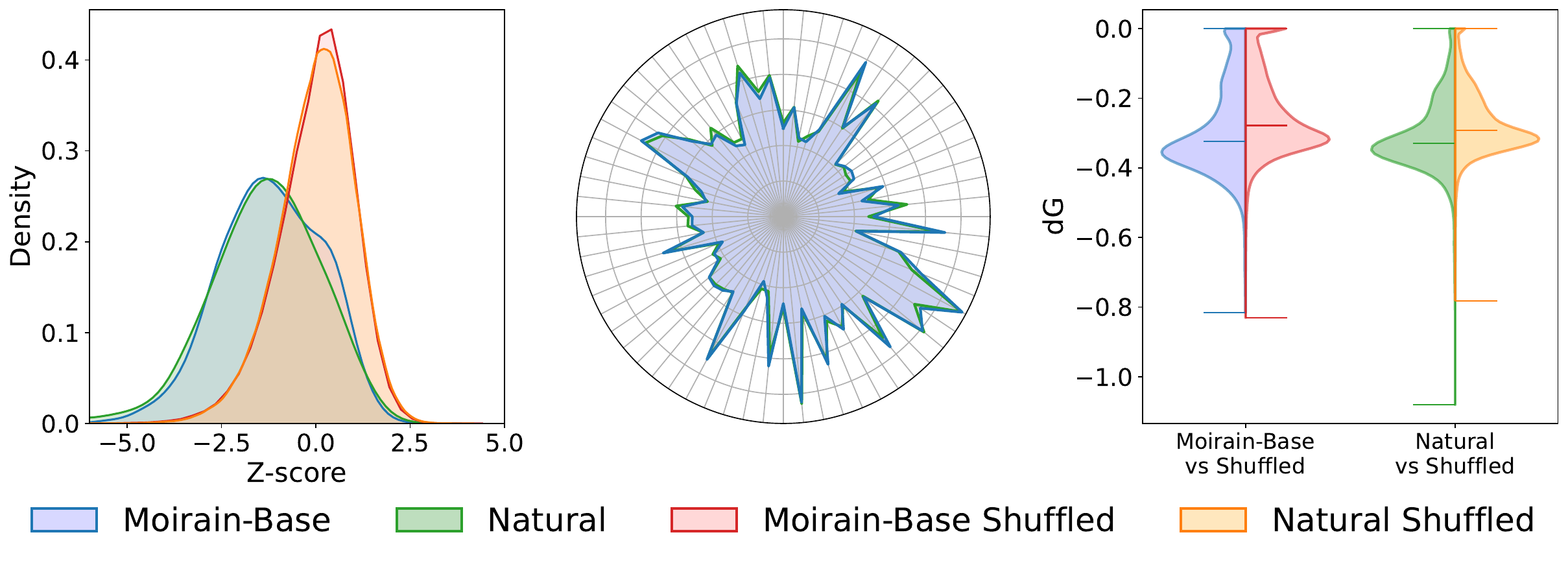}
  \caption{The comparison between generated (Moirain-Base) and natural RNA sequences. \textbf{(Left)} Z-score distributions for generated and natural sequences alongside their shuffled counterparts. \textbf{(Middle)} Radar plot of 3-mer frequency distributions (Fidelity $=0.018$), where each angular axis represents a distinct 3-mer. \textbf{(Right)} Distribution of dG shown via split-violin plots. For both the generated (left) and natural (right) groups, the left half of the violin represents the original sequences and the right half represents the shuffled controls. Horizontal markers indicate the median, minimum, and maximum values.}
  \label{fig:plaus_085}
\end{figure}

While the main text reports Moirain-Base performance under full random sampling, we also examined its behavior across different Top-p thresholds. By reducing this parameter to $p=0.85$, the 3-mer, dG, and Z-score distributions more closely resemble those of natural sequences (Figure~\ref{fig:plaus_085}). However, this gain in plausibility results in a trade-off, as Novelty falls to 0.24 and Diversity decreases to 0.50.

\begin{figure}
  \centering
  \includegraphics[width=\linewidth]{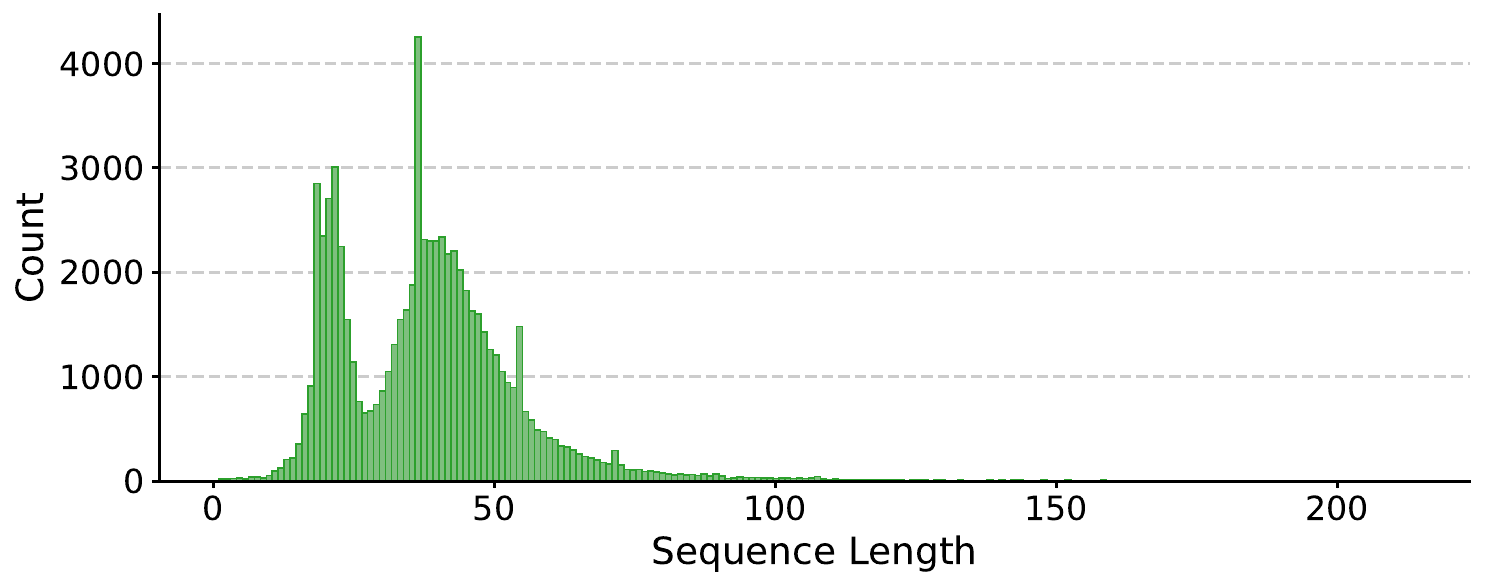}
  \caption{Sequence length distribution (nucleotides) of Moirain-DPO (following TOFU SFT) generations across the complete test set.}
  \label{fig:lengths}
\end{figure}

As illustrated in Figure~\ref{fig:lengths}, Moirain-DPO generates a broad distribution of sequence lengths, centered primarily between 40–45 nucleotides. Importantly, an additional sharp peak emerges at approximately 21-23 nucleotides, which matches the characteristic length of microRNAs. This suggests the model is successfully capturing specific biological archetypes within its broader length distribution.

Evaluation results on the "zero-similarity" subset are detailed in Table~\ref{tab:main_add}. While a performance drop is observed across all methods, RNA-BAnG exhibits the most resilience, displacing Moirain-DPO for the top overall ranking. Notably, TOFU SFT demonstrates superior generalization compared to CE, an advantage that becomes particularly evident following the subsequent DPO stage.

\begin{table}
  \caption{Binding affinity and authenticity benchmarking results for tested models. Performance is quantified by: (i) Hits$_x$ (0–1, $\uparrow$), representing target-specific binding success; (ii) Cov$_x$ (0–1, $\uparrow$), denoting the coverage of the target manifold; and (iii) Fidelity (0–1, $\downarrow$), measuring the proximity to natural interacting RNA.}
  \label{tab:main_add}
  \centering
  \sc
  \small
  \setlength{\tabcolsep}{5pt}
  \begin{tabular}{llcccccc}
    \toprule
    Method & Loss & Hits$_{0.5}$$\uparrow$ & Hits$_{0.7}$$\uparrow$ & Hits$_{0.9}$$\uparrow$ & Cov$_{0.25}$$\uparrow$ & Cov$_{0.5}$$\uparrow$ \\
    \midrule
    RNATranslator & & 0.40 & 0.31 & 0.21 & 0.42 & 0.25 \\
    RNA-BAnG & & \textbf{0.53} & \textbf{0.46} & \textbf{0.34} & \textbf{0.75} & \textbf{0.42} \\
    \midrule
    \multirow{2}{*}{Moirain-Multi} & CE & 0.41 & 0.31 & 0.19 & 0.50 & 0.25 \\
     & TOFU & 0.41 & 0.31 & 0.20 & 0.50 & 0.17  \\
    \multirow{2}{*}{Moirain-DPO} & CE & 0.36 & 0.27 & 0.15 & 0.50 & 0.08 \\
     & TOFU & 0.46 & 0.40 & 0.29 & 0.58 & 0.33 \\
    \bottomrule
  \end{tabular}
\end{table}

\section{Used Resources Licenses}
\label{app:licences}

This work utilizes a variety of open-access and proprietary resources. We employed datasets from RNAcentral Release 26 (CC0), RNAInter v4.0 (CC BY-NC 4.0), the RNA Compendium (CC BY 4.0), AlphaFold Protein Structure Database v6 (CC BY 4.0), and FoldSeek AFDB Cluster (version of 2025-09-12, CC BY 4.0). Computational analyses were performed using several software tools, including MMseqs2 (version 18-8cc5c, GNU General Public License v3.0), uShuffle (released Apr 20, 2020, custom free software license), BLASTP (version 2.12.0+, Public Domain), the ViennaRNA Package (custom free software license for research and education), and DeepCLIP (MIT). Additionally, we integrated structural predictions from AlphaFold 2 (Apache 2.0). All resources were used in accordance with their respective licensing terms.


\end{document}